\documentclass{article}
\usepackage{graphicx}
\usepackage{psfig}
\usepackage{epsfig}

\newcommand{\beq}{\begin{equation}}
\newcommand{\eeq}{\end{equation}}
\newcommand{\bea}{\begin{eqnarray}}
\newcommand{\eea}{\end{eqnarray}}

\newcommand{\Pom}{I\!P}

\newcommand{\FLUKA}{{\sc FLUKA}}

\newcommand{\DPMJET}{{\sc DPMJET}}
\def\dm2{\Delta m^2}
\def\sq2{sin^2(2\Theta)}

\begin{document}

\title{The FLUKA Monte Carlo, non--perturbative QCD and Cosmic Ray cascades}
\author{G.~Battistoni$^a$, A.~Fass\` o$^{b}$,A.~Ferrari$^{a,c}$, J.~Ranft$^d$ 
\\ 
P.~R.~Sala$^{a}$ } 
\date{	}
\maketitle
 
\begin{center}
$a)$ INFN Sezione di Milano, 20133 Milano, Italy \\
$b)$ SLAC, P.O. Box 4349, MS 48, Stanford, CA 94309, USA \\
$c)$ CERN, Geneva 23, Switzerland \\
$d)$ Fachbereich Physics, University of Siegen, 57068, Germany \\
\date{}
\maketitle
 
\end{center}
 
%
%
%
%
%

\begin{abstract}
The FLUKA Monte Carlo code, presently used in cosmic ray physics, contains
packages to sample soft hadronic processes which are built according to the
Dual Parton Model. This is a phenomenological model capable of reproducing
many of the features of hadronic collisions in the non
perturbative QCD regime. The basic principles of the model are summarized
and, as an example, the associated $\Lambda$--$K$ production is discussed.
This is a process which has some relevance for the calculation of 
atmospheric neutrino fluxes.
\end{abstract}

\section{Introduction}
\label{sec:intro}
The description of cosmic ray interactions in the Earth's atmosphere
requires a model which is capable of reproducing the dynamics of hadronic
interactions in a wide energy range, including phase space regions which
have not been experimentally accessed at accelerators. Therefore models
which are theoretically founded are the best candidates to cover reliably 
the whole region of interest.
Therefore QCD is the fundamental reference in the development of modern
codes for atmospheric showers. Unfortunately the bulk of hadronic
interactions belongs to the non perturbative regime of QCD, and therefore
we have not exactly calculable models that can be safely applied in cosmic
ray physics. In order to address the non perturbative region different
attempts have been devised, arriving to the definition of phenomenological
schemes with parameters that must be adjusted on the basis of experimental
data. The FLUKA Monte Carlo code\cite{fluka}, an all--purpose code for
transport and interaction of particles and nuclei already applied in cosmic
ray physics\cite{flukacr}, contains hadronic packages belonging to the
above defined 
class. Here we review the main features of the high energy hadron
interaction models of FLUKA. The specific features which bring to the
production of strange and charmed particles are then reviewed in more detail
in view of their specific importance in some aspects of cosmic ray physics.

\section{The high energy hadronic models in FLUKA}
\label{sec:hadmod}
\FLUKA{} is based, as far as possible, on the ``microscopic'' approach,
      {\it i.e.} the one starting from the basic hadron constituents
and from their known properties. Each step has to be
self--consistent and must be based on accepted physical bases.
Performances are optimized comparing with particle production data at
single interaction level. No tuning whatsoever is performed on ``integral''
      data, such as calorimeter resolutions, thick target yields, etc. 
Therefore, final predictions are obtained with a minimal set of free
      parameters, fixed for all energies and target/projectile combinations.
Results in complex cases as well as scaling laws and properties come forth
naturally from the underlying physical models and the basic conservation
laws are fulfilled {\it a priori}.

A comprehensive understanding of hadron--nucleon (h--N) interactions over a
wide energy range is of course a basic ingredient for a sound description of
hadron--nucleus ones.
Elastic, charge exchange and strangeness exchange reactions
are described as far as possible by phase--shift analysis and/or fits
of experimental differential data. Standard eikonal
approximations are often used at high energies.

At intermediate energies, the inelastic channel with the lowest threshold
(single pion production) opens
already around 290~MeV in nucleon-nucleon interactions,
and becomes important above 700~MeV. In pion-nucleon interactions
the production threshold is as low as 170~MeV.
Both reactions are normally described
in the framework of the isobar model: all reactions
proceed through an intermediate state containing at least one
resonance. There are two main classes of reactions, those which form a
resonant intermediate state (possible in $\pi$-nucleon reactions) and those
which contain two particles in the intermediate state. The former exhibit
bumps in the cross sections corresponding to the energy of the formed
resonance. Partial cross sections can be obtained from one--boson exchange
theories and/or folding of Breit--Wigner with matrix elements fixed by N--N
scattering or experimental data.
Resonance energies, widths, cross sections, and branching
ratios are extracted from data and conservation laws, whenever possible,
making explicit use of spin and isospin relations.
They can be also inferred from inclusive cross sections when needed.
For a discussion of resonance production, see for
example~\cite{RES1,RES2,RES3}.

As soon as the projectile energy exceeds a few GeV, the description in
terms of resonance production and decay becomes more and more
difficult. The number of resonances which should be considered grows
exponentially and their properties are often poorly known. Furthermore, the
assumption of one or two resonance creation is unable to reproduce the
experimental finding that most of the particle production at high energies
occurs neither in the projectile nor in the target fragmentation region,
but rather in the central region, for small values of Feynman~$x$
variable. Different models, based directly on quark degrees of freedom,
must be introduced. 

The features of ``soft''
interactions (low-$p_T$ interactions) cannot be derived from the QCD
Lagrangian, because the large value taken by the
running coupling constant prevents the use of perturbation theory.
Models based on interacting strings have emerged as a powerful tool in
understanding QCD at the soft hadronic scale, that is in the
non-perturbative regime. 

A theory of interacting strings can be managed by means of 
the Reggeon-Pomeron calculus in the 
framework of perturbative Reggeon Field Theory\cite{Collins},
an expansion already developed before the 
establishment of QCD.
Regge theory makes use explicitly of the constraints of analyticity and 
duality. 
On the basis of these concepts, calculable models can be constructed
and one of the most successful attempts in this field is the so called 
``Dual Parton Model'' (DPM), originally developed in Orsay in
1979~\cite{DPMORI}.  
It provides the theoretical framework 
to describe hadron-nucleon interaction from several
GeV onwards. 
In DPM a hadron is a low-lying excitation of an open
string with quarks, antiquarks or diquarks sitting at its ends. In
particular mesons  are
described as strings with their valence quark and antiquark at the
ends. (Anti)baryons are treated like open strings with a (anti)quark and a
(anti)diquark at the ends, made up with their valence quarks.

\begin{figure}[hbtp]
\centerline{\epsfxsize=8.cm\epsfbox{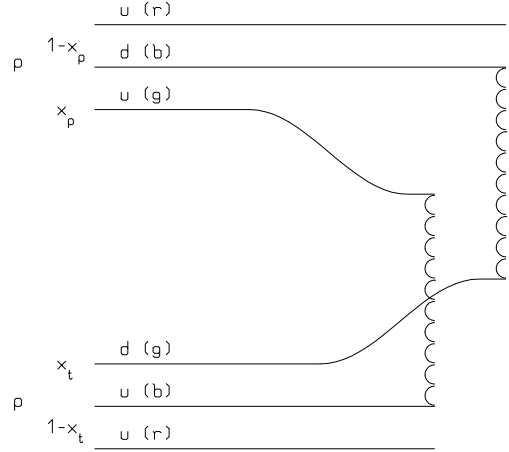}}   
\caption{Leading two-chain diagram in DPM for $p-p$ scattering. The color
(red, blue, and green) and quark combination shown in the figure is just 
one of the allowed possibilities.\label{fig:ppchain} }
\end{figure}

At sufficiently high energies,
the leading term in high energy scattering 
corresponds to a ``Pomeron'' ($\Pom$) exchange (a closed string exchange
with the quantum numbers of vacuum), 
which has a cylinder topology. By means of the optical theorem, connecting
the forward elastic scattering amplitude to the total inelastic cross
section, it can be shown that 
from the Pomeron topology it follows that two hadronic 
chains are left as the sources of particle production
(unitarity cut of the Pomeron). 
While the partons 
(quarks or diquarks) out of which chains are stretched carry a net 
color, the chains themselves are built in such a way to carry no net 
color, or to be more exact to constitute color singlets like all 
naturally occuring hadrons. In practice, as a consequence of color
exchange in the interaction, each colliding hadron splits into two
colored system, one carrying color charge $c$ and the other $\bar c$.
These two systems carry together the whole momentum of the hadron. The
system with color charge $c$ ($\bar c$) of one hadron combines with the
system of complementary color of the other hadron, in such a way to
form two color neutral chains. These chains appear as two back-to-back
jets in their own centre-of-mass systems.
The exact way of building up these chains depends on the nature of the
projectile-target combination (baryon-baryon, meson-baryon,
antibaryon-baryon, meson-meson). Let us take as example the
case of nucleon-nucleon (baryon-baryon) scattering.
In this case, indicating with $q^v_p$ the valence quarks of the projectile, 
and with $q^v_t$ those of the target, and assuming that the quarks sitting 
at one end of the baryon strings carry momentum fraction 
$x^v_p$ and $x^v_t$ respectively, the resulting chains are
$q^v_t-q^v_p q^v_p$ and $q^v_p-q^v_t q^v_t$, as shown in
fig.~\ref{fig:ppchain}.

Energy and momentum in the centre-of-mass system of 
the collision, as well as the invariant mass squared of 
the two chains, can be obtained from:
\bea
E^*_{ch1} & \approx & \frac{\sqrt{s}}{2} ( 1 - x^v_p + x^v_t ) \nonumber \\
E^*_{ch2} & \approx & \frac{\sqrt{s}}{2} ( 1 - x^v_t + x^v_p ) \nonumber \\
p^*_{ch1} & \approx & \frac{\sqrt{s}}{2} ( 1 - x^v_p - x^v_t ) =
- p^*_{ch2} \label{eq:xeps} \\
s_{ch1} & \approx & s ( 1 - x^v_p ) x^v_t \nonumber \\
s_{ch2} & \approx & s ( 1 - x^v_t ) x^v_p \nonumber
\eea

The single Pomeron exchange diagram is the dominant contribution,
however higher order contributions with multi-Pomeron exchanges become
important at energies in excess of 1~TeV in the laboratory. They
correspond to more complicated topologies, and DPM provides a way for
evaluating the weight of each, keeping into account the unitarity
constraint. Every extra Pomeron exchanged gives rise to two
extra chains which are built using two $q\bar q$ couples excited from
the projectile and target hadron sea respectively. The inclusion of 
these higher order diagrams is usually referred  to as {\it multiple 
soft collisions}.

Two more ingredients are required to completely settle the problem. The 
former is the momentum distribution for the $x$ variables of valence 
and sea quarks. Despite the exact form of the momentum distribution
function, $P(x_1,..,x_n)$, is not known, general considerations based on
Regge arguments allow to predict the asymptotic behavior of this 
distribution whenever each of its arguments goes to zero. The behavior 
turns out to be singular in all cases, but for the diquarks. A 
reasonable assumption, always made in practice, is therefore to 
approximate the true unknown distribution function with the product of 
all these asymptotic behaviors, treating all the rest as a 
normalization constant.

The latter ingredient is a hadronization model, which must 
take care of transforming each chain into a sequence of physical 
hadrons, stable ones or resonances. There are two basic assumptions.
One is that of
{\it chain universality}, which assumes that once the chain ends and the 
invariant mass of the chain are given, the hadronization properties are 
the same regardless of the physical process which originated the chain.
Therefore the knowledge coming from hard processes and $e^+e^-$ 
collisions about hadronization can be used to fulfill this task.
The other is the scale invariance of fragmentation functions, when
$\sqrt{s}$ is much larger than the masses.
There 
are many more or less phenomenological models which have been developed 
to describe hadronization (examples can be found 
in~\cite{JETSET,BAMJET}). In principle hadronization properties too can 
be derived from Regge formalism~\cite{Kaifrag1}. The original \FLUKA{}
model makes use of the hadronization model of \cite{BAMJET}.

It is possible to extend DPM to hadron-nucleus collisions 
too~\cite{DPMORI}, making use of the so called Glauber-Gribov approach. 
The Glauber-Gribov model~\cite{GRIBOV,GRIBOV2,GRIBOV3} represents the
diagram interpretation of the Glauber cascade.
The $\nu$ interactions of the projectile originate
2$\nu$ chains, out of which
2 chains (valence-valence chains) struck between the projectile and
target valence (di)quarks,
$2(\nu-1)$ chains (sea-valence chains) between projectile sea $q-\bar q$
and target valence (di)quarks.

\begin{figure}[tbh]
\centerline{\epsfxsize=9cm\epsfbox{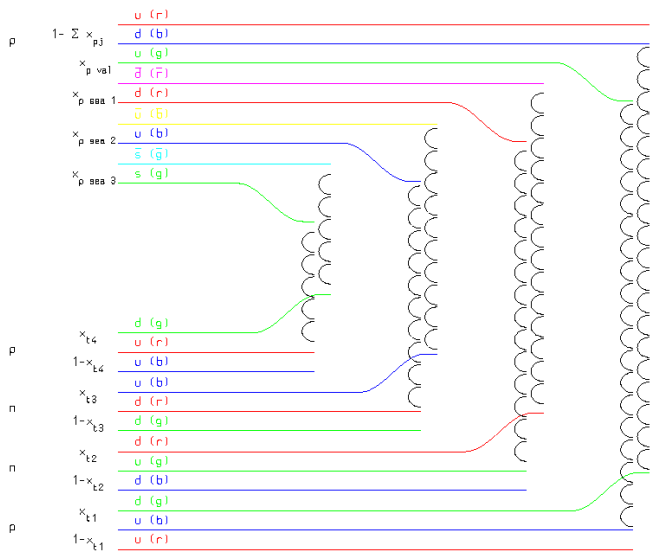}}   
\caption{
Leading two-chain diagrams in DPM for $p-A$ Glauber scattering with 4
collisions. The colour and quark combinations shown
in the figure are just one of the allowed possibilities.
\label{fig:pglauchain}
}
\end{figure}

A pictorial example of the chain building process is depicted in
fig.~\ref{fig:pglauchain} for p--A: similar diagrams apply to $\pi$--A and
$\bar p$ --A respectively.

The distribution of the projectile energy among many chains naturally
softens the energy distributions of reaction products
and boosts the multiplicity with respect to hadron-hadron interactions.
In this way, the model accounts for the major A-dependent features
without any degree of freedom, except in the treatment of mass effects at
low energies.

The Fermi motion of the target nucleons must be included to obtain the
correct kinematics, in particular the smearing of p$_T$ distributions.
All nuclear effects on the secondaries are accounted for by the subsequent
steps. In FLUKA these are performed in the framework of a generalized
intranuclear cascade (GINC) which embeds quantum correction effects like
for instance the {\it formation zone} concept, a sort of
``materialization" time. For further details see \cite{Trieste}.

At very high energies, those of interest for high energy cosmic ray studies 
(10--10$^5$~TeV in the lab), hard processes cannot be longer 
ignored. They are calculable by means of perturbative QCD and
can be included in DPM through proper unitarization 
schemes which consistently treat soft and hard processes together.
This is for instance what is implemented in the framework of
the ``2-component Dual Parton Model'' ({\it i.e.} soft + hard collisions)
in the \DPMJET{} model\cite{Ranft}. Here the hard collisions are those
which are described by the tree--level QCD diagrams.

The latest releases of \FLUKA{} are now interfaced to \DPMJET-II.53, 
so to allow the sampling of high energy nucleus-nucleus collisions for
E$_{lab}$ from 5-10~GeV/n up to 10$^{9}$-10$^{11}$ GeV/n).

The original interface to the \DPMJET-II.53 version has
 recently been upgraded to comply with the  \DPMJET-III version. Beyond the
full description of nucleus--nucleus collisions, the interface to
 \DPMJET{} allows to extend the \FLUKA{} energy limits 
 for hadronic simulations in general. 

\section{Strangeness production in high energy collisions in
  FLUKA} 
\label{sec:strange}
The production of particles containing {\it s} or heavier quarks in
nucleon--nucleon collisions 
requires to pick up such quarks from the sea. This, in the framework of
Dual Parton Model may occur for soft processes in two cases:
\begin{enumerate}
\item in the process of chain formation when valence and sea (di)quarks
  combine together.
This occurs in multi-Pomeron exchange (relevant only
  at high energy) and in the multiple chains deriving from the Glauber expansion in
  the case of nucleons in a nucleus.
\item in the process of chain hadronization, when the color string tension
  materializes in $q-\bar{q}$ or $qq-\bar{q}\bar{q}$ pairs. Numerically
  this is the most important contribution.
\end{enumerate}

In particular, the hadronization of valence diquark-quark chains is the
fundamental mechanism to produce K mesons and strange baryons. A reliable
description of this production can be relevant in cosmic ray
applications ons. In particular it has been pointed out how, for kinematical
reasons, kaons have a fundamental role in the generation of atmospheric
neutrinos at high energy\cite{gaisserk}, and the uncertainties on kaon production are one
of the important contributions to the systematic error in the calculations
of fluxes, especially above 100 GeV.

Table \ref{tab:tab1} shows the average multiplicity of some particles originating
from the hadronization of chains  
with invariant mass of 10 GeV occurring in p--nucleon scattering 
as obtained from \FLUKA{}. This table is calculated after that
e.m. and strong decays have already occurred (therefore there are no more
$\eta$'s, $K^\star$'s etc.).

\begin{table}[ht]
\begin{center}
\begin{tabular}{@{}lcccc@{}}
\hline
{} &{} &{} &{} &{}\\[-1.5ex]
{} & uu--u & uu--d & ud--u & ud--d\\[1ex]
\hline
{} &{} &{} &{} &{}\\[-1.5ex]
$p$ & 0.803 &  0.777 &  0.538 &  0.511 \\[1ex]
$\overline{p}$ & 0.073 &  0.073 &  0.074 &  0.074 \\[1ex]
$n$ & 0.241 &  0.265 &  0.511 &  0.538 \\[1ex]
$\overline{n}$ & 0.075 &  0.073 &  0.075 &  0.074 \\[1ex]
$\pi^+$ & 2.957 &  2.490 &  2.530 &  2.047 \\[1ex]
$\pi^-$ & 2.577 &  2.580 &  2.736 &  2.522 \\[1ex]
$\pi^0$ & 1.788 &  2.271 &  2.056 &  2.739 \\[1ex]
$K^+$ & 0.233 &  0.213 &  0.230 &  0.209 \\[1ex]
$K^0$ & 0.199 &  0.217 &  0.205 &  0.224 \\[1ex]
$K^-$ & 0.165 &  0.258 &  0.168 &  0.162 \\[1ex]
$\overline{K^0}$ & 0.160 &  0.159 &  0.163 &  0.163 \\[1ex]
$\Lambda$ & 0.064 &  0.064 &  0.066 &  0.066 \\[1ex]
$\overline{\Lambda}$ & 0.012 &  0.012 &  0.012 &  0.013 \\[1ex]
\hline
\end{tabular}\label{tab:tab1}
\caption{Particle multiplicities from the hadronization of chains 
with invariant mass of 10 GeV occurring in p--nucleon scattering 
as obtained from \FLUKA{}}
\end{center}
\end{table}

\begin{figure}[hbtp]
\begin{center}
\begin{tabular}{cc}
\includegraphics[height=4.cm]{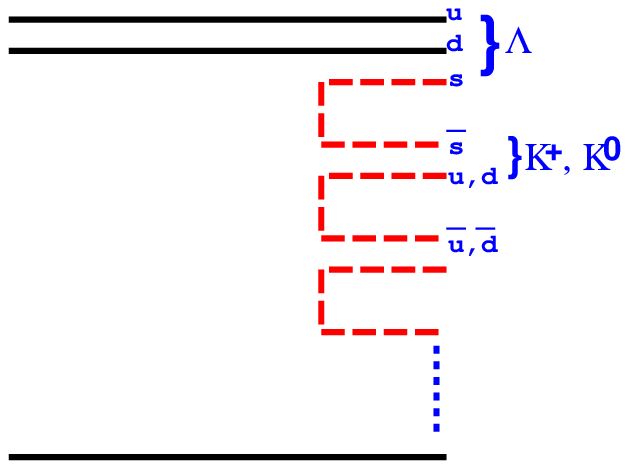} &
\includegraphics[height=4.cm]{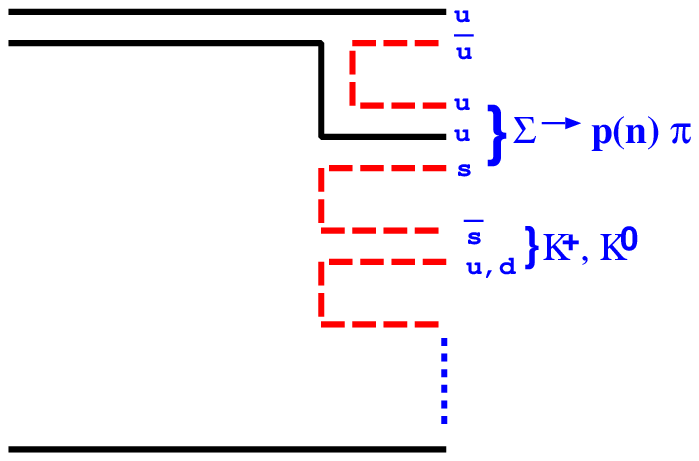} \\
\end{tabular}
\caption{\label{fig:had1}Left: associated $\Lambda$ $K$ production. Right:
  $\Sigma$ $K$ production with ``popcorn'' effect. }
\end{center}
\end{figure}

\begin{figure}[hbtp]
\begin{center}
\begin{tabular}{cc}
\includegraphics[height=4.cm]{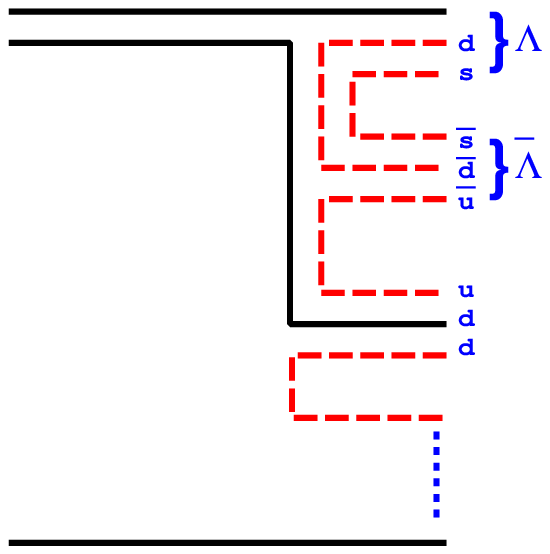} &
\includegraphics[height=4.cm]{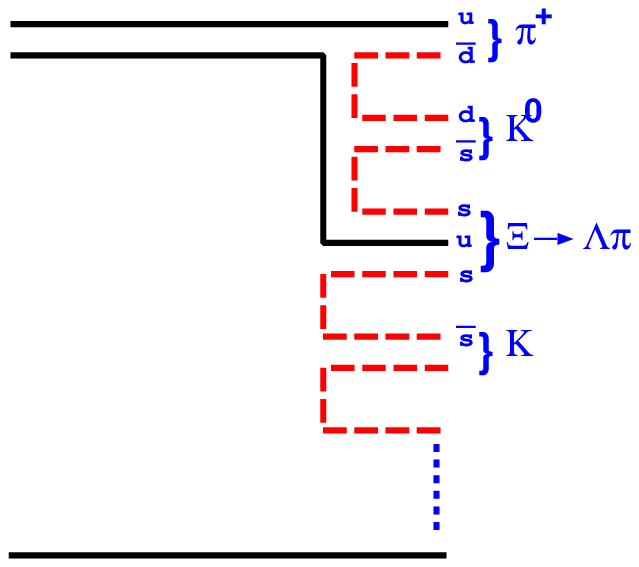} \\
\end{tabular}
\caption{\label{fig:had2} Left: $\Lambda$ $\bar \Lambda$ production with
  double ``popcorn'' effect. Right: Another example of ``popcorn'' leading
  to $\Xi$ $K$ production }
\end{center}
\end{figure}

\begin{figure}[hbtp]
\begin{center}
\includegraphics[height=4.cm]{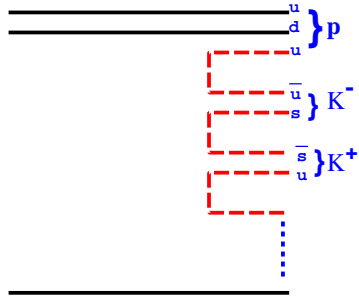}
\caption{\label{fig:had3}Example where $K^+$ $K^-$ (or $K^0$ $\bar {K^0}$
  if $d$ quark replace the $u$ one) are produced, without any associated
  strange baryon. }
\end{center}
\end{figure}

The extraction of $s--\bar s$ pairs in the hadronization process may give
rise to different configurations. Some examples are reported in
Figg.\ref{fig:had1},\ref{fig:had2},\ref{fig:had3}, where we consider the
string between a valence diquark from the projectile and a valence quark from the
target (solid lines). The dashed lines represent the $q$--$\bar q$ or $qq$--$\bar
{q}\bar{q}$ pairs materialized in the process. It must be intended that
with a proper probability the suggested particles are replaced by the
corresponding excited states (resonances) like $K^\star$'s.

As far as particle production in cosmic ray showers is concerned, an
interesting case is 
the production at high rapidity (or $x_{F}$). On average, the valence
diquark from the projectile in the laboratory frame carries the highest
fraction of momentum of the incident nucleon. 

Therefore, on average, the 
the particles which are produced close in rapidity to the projectile
diquark give a major contribution to the yield (see for instance the
concept of ``spectrum weighted moment''\cite{mom}). The case shown in the
left  panel of Fig.\ref{fig:had1} is the typical case in which the
excitation of a $s--\bar s$ pair leads to the associated $\Lambda$ $K$
production. The valence diquark can dissociate before hadronization (using
the mechanism that has been called ``popcorn'' in the Lund chain
fragmentation model JETSET\cite{popcorn}), as shown in the
right panel of Fig.\ref{fig:had1} and in Fig.\ref{fig:had2}. Sometimes
a strange baryon might not be associated to a $K$ meson, but to the
corresponding anti--baryon, as shown in the left panel of
Fig.\ref{fig:had2}. In other cases, $K$ meson can be produced in pairs without the
association to a strange baryon. An example is shown in Fig.\ref{fig:had3}.

It can be expected that $K$ mesons associated to strange baryons 
might be harder, in average, than those who are not associated. This because
the baryons have a higher probability to be produced closer in rapidity to
the diquark from the projectile. In the case of $\Lambda$, which contains
an $s$ valence quark, this feature must be visible for the associated $K^+$
(and $K^0$), containing an $\bar s$ valence quark, and not for $K-$.

\begin{figure}[thb]
\begin{center}
\centerline{\epsfxsize=12.cm\epsfbox{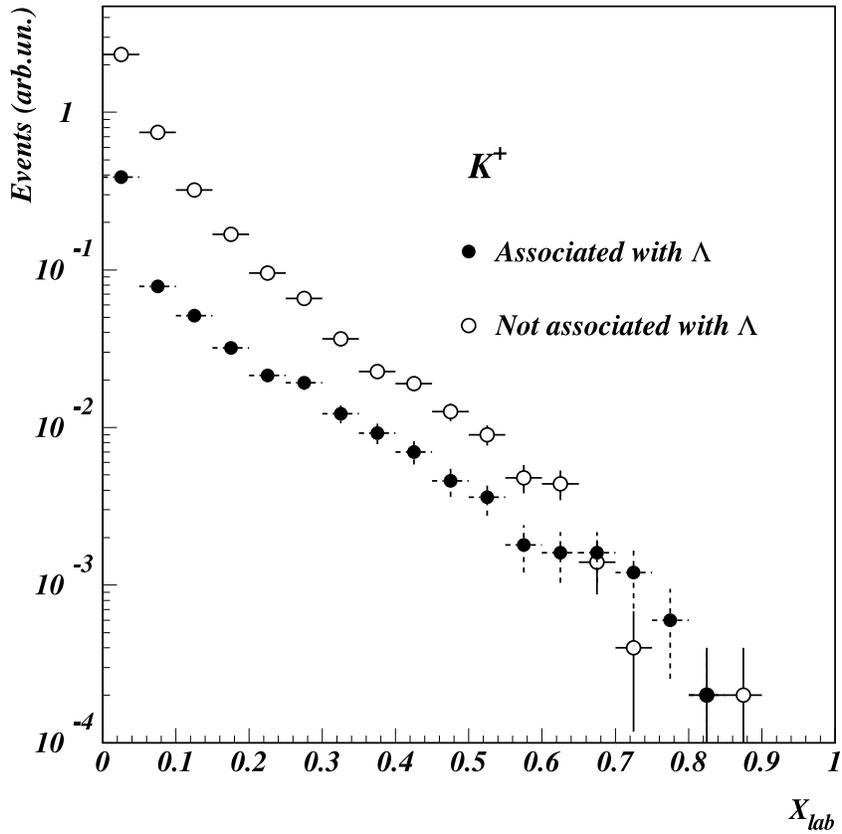}}   
\vspace*{-2cm}
\caption{\label{fig:kz1} $X_{lab}$ distributions of $K^+$ 
produced in p--Air collisions at $E_0$ = 100 GeV. Events where
  $K$ mesons are associated to $\Lambda$ production are distinguished from
  the others.}
\end{center}
\end{figure}

\begin{figure}[p]
\begin{center}
\centerline{\epsfxsize=12.cm\epsfbox{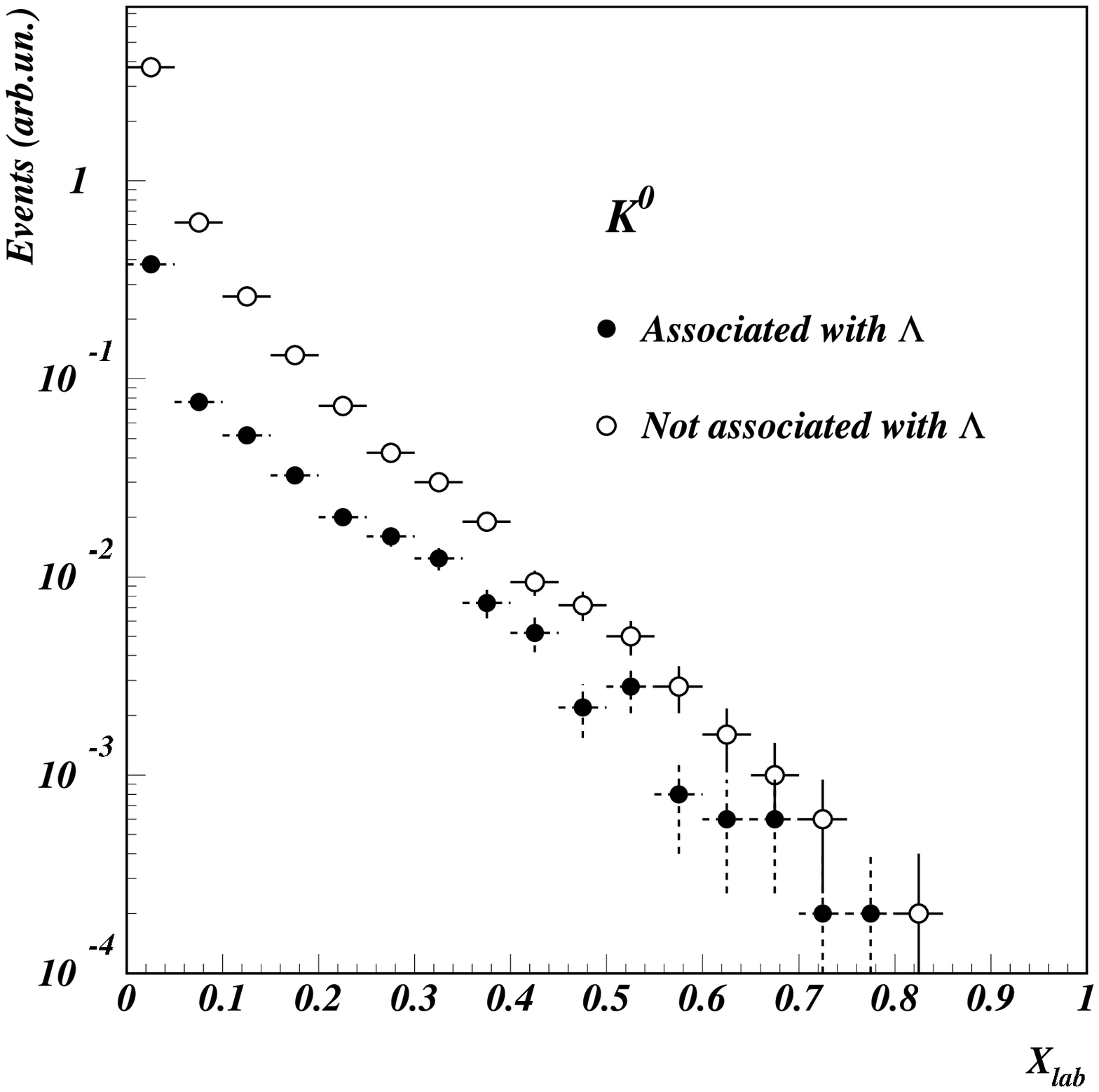}}   
\vspace{-2cm}
\caption{\label{fig:kz2} $X_{lab}$ distributions of $K^0$
produced in p--Air collisions at $E_0$ = 100 GeV. Events where
  $K$ mesons are associated to $\Lambda$ production are distinguished from
  the others.}
\end{center}
\end{figure}

\begin{figure}[p]
\begin{center}
\centerline{\epsfxsize=12.cm\epsfbox{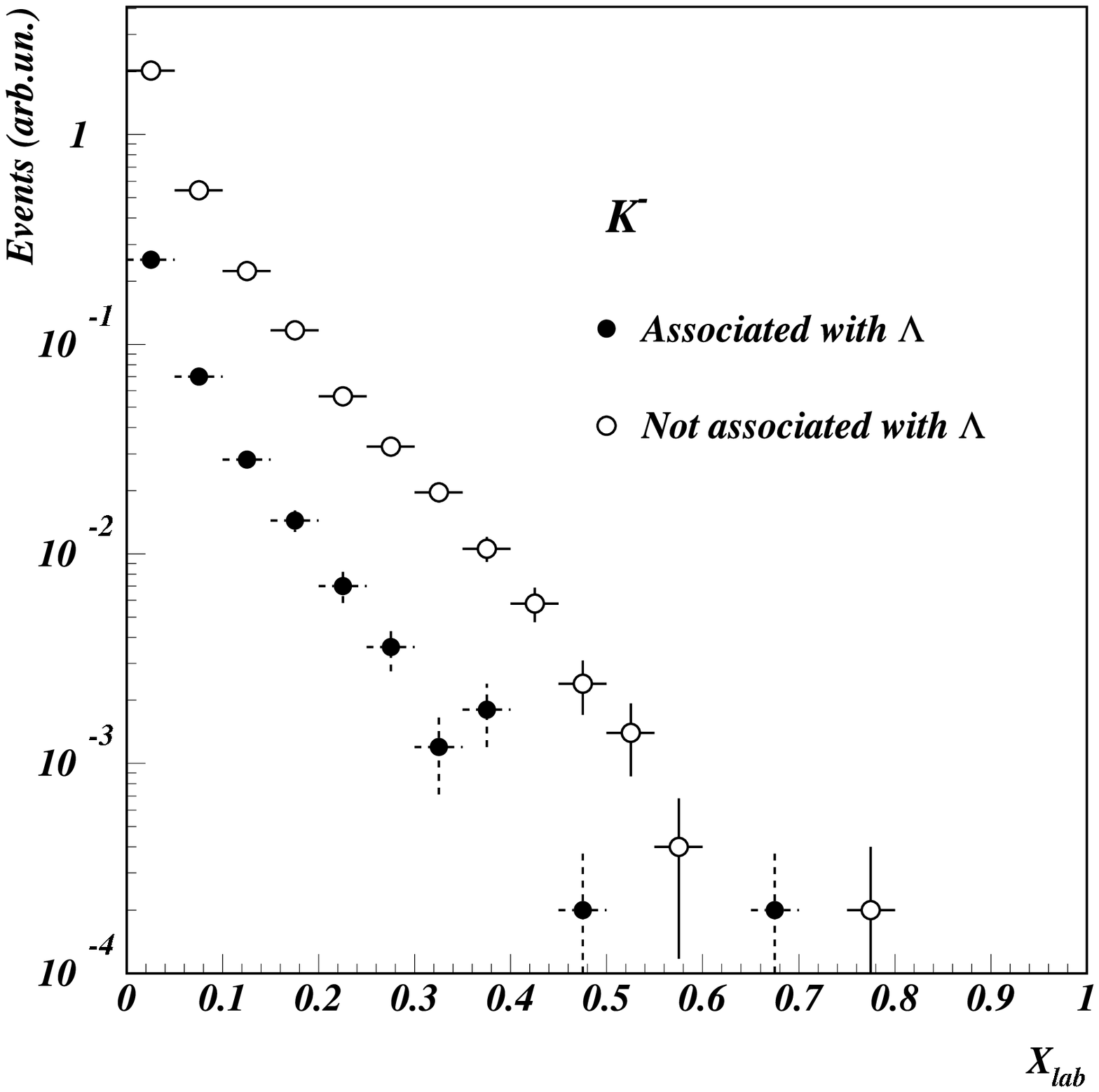}}   
\vspace{-2cm}
\caption{\label{fig:kz3} $X_{lab}$ distributions of $K^-$
produced in p--Air collisions at $E_0$ = 100 GeV. Events where
  $K$ mesons are associated to $\Lambda$ production are distinguished from
  the others.}
\end{center}
\end{figure}

In order to test this prediction we have generated p--Air interactions at
analyzing the distribution of $X_{lab}$ = $E/E_0$ of the highest rapidity
$K$ meson, distinguishing events in 
which $\Lambda$ were produced from the rest. The results for $E_0$ = 100
GeV are shown in Fig.\ref{fig:kz1} for $K^+$, in Fig.\ref{fig:kz2} for $K^0$ and in
Fig.\ref{fig:kz3} for $K^-$. These last, together with $\bar K^0$, are
mostly associated to the correspondent anti--meson.

The expectation for a harder spectrum of $\Lambda$ associated $K^+$ and
 $K^0$ is confirmed. However the fraction of events with strange baryons
 production is only $\sim$0.12 at this energy. 

\begin{figure}[p]
\begin{center}
\includegraphics[width=12.cm]{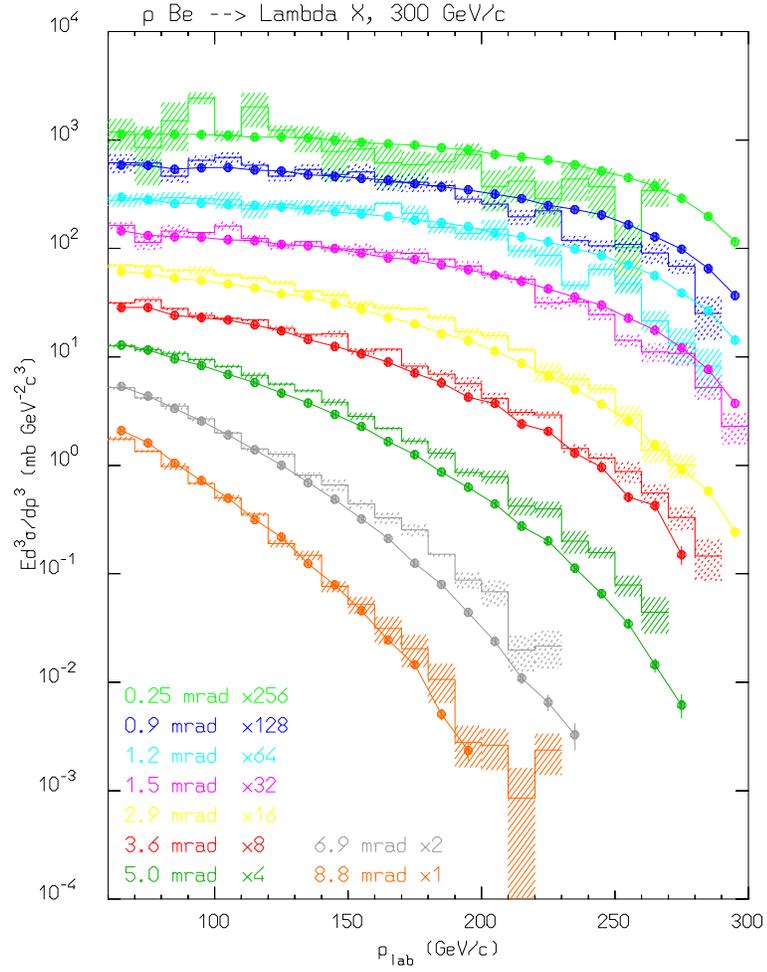}
\caption{\label{fig:lam300} Double differential cross section in laboratory
  momentum and angle for 
$\Lambda$ produced in p--p interaction at 300 GeV in the laboratory frame
Points are experimental data \protect\cite{lam300c}) and shaded histogram
is the \FLUKA{} Monte Carlo prediction}
\end{center}
\end{figure}

The capability of \FLUKA{} of reproducing experimental data on $\Lambda$
production, at least for few hundreds of GeV of proton energy, is shown in
Fig.\ref{fig:lam300}, where the measured 
double differential cross section in laboratory momentum and angle for
$\Lambda$ produced in p--p interaction at 300 GeV in the laboratory frame
(points\cite{lam300c}) are compared to Monte Carlo
predictions (shaded histograms). The typical hard spectrum of forward
produced baryons is clearly reproduced in shape and in absolute value.

Analogous mechanisms apply for charmed particle production. However
the transverse mass suppression is quite strong in the Lund string fragmentation
model, leading to a probability of creating a $c$--$\bar c$ pair of the
order of $u~\bar u$ : $d~\bar d$ : $s~\bar s$ : $c~\bar c$ = 1 : 1 : 0.3 :
10$^{-11}$\cite{JETSET}. In the BAMJET model\cite{BAMJET} this probability
is higher
but enough to be significant ($\sim$ 10$^{-5}$). 
In case of multiple collisions, as in nucleon--nucleus scattering, or for
multi--Pomeron exchange, where color chains are built also with quarks
and anti--quarks from the sea sitting at one or both ends, 
additional heavy quark production is possible.
In any case, at very
high energies, it has been shown that the dominant contribution in charm
production comes from 
the hard parton--parton collisions as described by perturbative
QCD\cite{dpmcharm}. For these processes in this energy range \FLUKA{}
relies on the interface to \DPMJET.

\section*{Acknowledgments}
The authors wish to thank the organizer of the workshop on ``QCD at cosmic
ray energies'' and  T.K.~Gaisser for having stimulated this work.
This work has been supported in part by DOE (contract no. 
DE-AC02-76SF00515).


\end{document}